\documentclass[12pt,preprint]{aastex}
\newcommand{\NHunits}{\mbox{$\times 10^{21} {\rm cm}^{-2}$}}

\newcommand{\xrtposra}{\mbox{RA(J2000)\,=\,23$^{\rm h}$53$^{\rm m}$53$\fs0$}}
\newcommand{\xrtposdec}{\mbox{Dec(J2000)=$+65\degr$56$\arcmin$19\farcs8}}
\newcommand{\xrtflux}{\mbox{$1.1\times 10^{-8}$ergs cm$^{-2}$ s$^{-1}$}}
\newcommand{\xrtfluxerr}{\mbox{$(1.1\pm0.3)\times 10^{-8}$ergs cm$^{-2}$ s$^{-1}$}}

\begin{document}
\title{GRB 050117: Simultaneous Gamma-ray and X-ray Observations with the {\it \bfseries Swift} Satellite}
\author{Joanne E. Hill\altaffilmark{1,2}, David C. Morris\altaffilmark{3}, Takanori~Sakamoto\altaffilmark{1}, Goro~Sato\altaffilmark{4}, David N. Burrows\altaffilmark{3}, Lorella Angelini\altaffilmark{1,5}, Claudio~Pagani\altaffilmark{3,6}, Alberto~Moretti\altaffilmark{6},  Antony~F.~Abbey\altaffilmark{7}, Scott~Barthelmy\altaffilmark{1}, Andrew~P.~Beardmore\altaffilmark{8},  Vadim~V.~Biryukov\altaffilmark{9}, Sergio~Campana\altaffilmark{6}, Milvia~Capalbi\altaffilmark{10}, Giancarlo~Cusumano\altaffilmark{11}, Paolo~Giommi\altaffilmark{10}, Mansur~A.~Ibrahimov\altaffilmark{12}, Jamie~Kennea\altaffilmark{3}, Shiho~Kobayashi\altaffilmark{3,13}, Kunihito~Ioka\altaffilmark{3, 13}, Craig~Markwardt\altaffilmark{1}, Peter~M\'{e}sz\'{a}ros\altaffilmark{3}, Paul~T.~O'Brien\altaffilmark{8}, Julian~P.~Osborne\altaffilmark{8}, Alexei~S.~Pozanenko\altaffilmark{14}, Matteo~Perri\altaffilmark{10}, Vasilij~V.~Rumyantsev\altaffilmark{15}, Patricia~Shady\altaffilmark{16}, Dmitri~A.~Sharapov\altaffilmark{12}, Gianpiero~Tagliaferri\altaffilmark{6}, Bing~Zhang\altaffilmark{17}, Guido~Chincarini\altaffilmark{6, 18}, Neil~Gehrels \altaffilmark{1}, Alan~Wells\altaffilmark{3,7}, John~A.~Nousek\altaffilmark{3}}

\altaffiltext{1}{NASA/Goddard Space Flight Center, Greenbelt, MD 20771, USA; {\it jhill@milkyway.gsfc.nasa.gov}}
\altaffiltext{2}{Universities Space Research Association, 10211 Wincopin Circle, Suite 500, Columbia, MD, 21044-3432, USA}
\altaffiltext{3}{Department of Astronomy \& Astrophysics, 525 Davey
  Lab., Pennsylvania State
	University, University Park, PA 16802, USA}
\altaffiltext{4}{ Institute of Space and Astronautical Science/Japan Aerospace Exploration Agency, Bldg. A/Room 1453, 3-1-1 Yoshinodai, Sagamihara, Kanagawa, 229-8510 Japan}
\altaffiltext{5}{Department of Physics and Astronomy, The Johns Hopkins University, 3400 North Charles Street, Baltimore, MD 21218, USA }
\altaffiltext{6}{INAF -- Osservatorio Astronomico di Brera, Via Bianchi 46, 23807 Merate, Italy}
\altaffiltext{7}{Space Research Centre, University of Leicester, Leicester LE1 7RH, UK}
\altaffiltext{8}{Department of Physics and Astronomy, University of Leicester, Leicester LE1 7RH, UK}
\altaffiltext{9}{Crimean Laboratory of Sternberg Astronomical Institute, Moscow
University, Russia}
\altaffiltext{10}{ASI Science Data Center, via Galileo Galilei, 00044 Frascati, Italy}
\altaffiltext{11}{INAF- Istituto di Astrofisica Spaziale e Fisica Cosmica Sezione di Palermo,  
                 Via Ugo La Malfa 153, 90146 Palermo, Italy}
\altaffiltext{12}{Ulugh Beg Astronomical Institute, Tashkent 700052, Uzbekistan}
\altaffiltext{13}{Center for Gravitational Wave Physics, Pennsylvania State University, University Park, PA 16802, USA}
\altaffiltext{14}{Space Research Institute, Moscow 117810, Russia}

\altaffiltext{15}{Crimean Astrophysical Observatory, Ukraine}
\altaffiltext{16}{Mullard Space Science Laboratory, Holmbury St. Mary, Dorking, Surrey, UK}
\altaffiltext{17}{Department of Physics, University of Nevada, Box 454002, Las Vegas, NV 89154-4002, USA}
\altaffiltext{18}{Universit\`a degli studi di Milano-Bicocca,
                 Dipartimento di Fisica, Piazza delle Scienze 3, I-20126 Milan, Italy}

\begin{abstract}
The {\it Swift} Gamma-Ray Burst Explorer performed its first autonomous, X-ray follow-up to a newly detected GRB on 2005 January 17, within 193 seconds of the burst trigger by the {\it Swift} Burst Alert Telescope. While the burst was still in progress, the X-ray Telescope obtained a position and an image for an un-catalogued X-ray source; simultaneous with the gamma-ray observation. The XRT observed flux during the prompt emission was \xrtflux \ in the 0.5--10 keV energy band. The emission in the X-ray band decreased by three orders of magnitude within 700 seconds, following the prompt emission. This is found to be consistent with the gamma-ray decay when extrapolated into the XRT energy band. During the following 6.3 hours, the XRT observed the afterglow in an automated sequence for an additional 947 seconds, until the burst became fully obscured by the Earth limb. A faint, extremely slowly decaying afterglow, $\alpha=-0.21$, was detected. Finally, a break in the lightcurve occurred and the flux decayed with $\alpha<-1.2$. The X-ray position triggered many follow-up observations: no optical afterglow could be confirmed, although a candidate was identified 3 arcsecs from the XRT position.
\end{abstract}   


\section{Introduction}

The {\it Swift} Gamma-ray Burst Explorer \citep{Gehrels} was launched on 2004 November 20 to study gamma-ray bursts (GRBs) over a broad energy band covered by three instruments; a wide field of view gamma-ray Burst Alert Telescope (BAT) and two Narrow Field Instruments (NFIs); an X-ray Telescope (XRT) and a UV--Optical Telescope (UVOT). {\it Swift} slews rapidly to bursts detected by the Burst Alert Telescope (BAT) \citep{Barthelmy}, in order to point the XRT \citep{Burrows} and UVOT \citep{Roming} telescopes at the burst for rapid follow-up observations of the afterglow. 

On 2005 January 17 at 12:52:36.037 UT, the {\it Swift} Burst Alert Telescope triggered and located GRB 050117 \citep{GCN 2952}. For the first time, {\it Swift} responded autonomously to the BAT triggered burst, pointing the XRT at the GRB while the burst was still in progress, and allowing simultaneous gamma-ray and X-ray flux measurements of the prompt emission and follow up observations of the afterglow. The BAT lightcurve of the burst, which lasted 220 seconds, is multi-peaked (Figure \ref{f_energy_res_lc}). The XRT was on target and obtained a refined position and an image within 193 seconds of the BAT detection. The XRT detected the GRB at the end of the burst on-set, measuring a source position of $\xrtposra$  $\xrtposdec$  \citep{GCN 2955} and an absorbed flux of $\xrtflux$ in the 0.5--10 keV band. A faint afterglow was detected by the XRT during the subsequent orbits. The UVOT activation was not complete at the time of these observations and therefore it remained in a non-observing state throughout. No radio or optical afterglow was detected by the ground based follow-up observations.

We report on the first autonomous re-pointing of an X-ray Telescope to a newly discovered GRB, and describe the transition from the prompt emission to the afterglow from the resulting lightcurve.

\section{Gamma-ray Observations}
Ground analysis of the BAT Ôevent-by-eventÕ data produces an image significance of 54.2 sigma and a refined GRB position within 10.2 arcsecs of the XRT position of \linebreak RA(J2000)=23$^{\rm h}$53$^{\rm m}$53$\fs1$ Dec(J2000)=$+65\degr$56$\arcmin$10\farcs3, with an error circle radius of 45.6 arcsec at 90\% confidence level. Figure \ref{f_image} shows the BAT error circle super-imposed on the XRT image.

The burst prompt emission lasted 220 seconds, with $T_{90}$ duration of $167.896\pm0.004$ seconds. This indicates that GRB 050117 is a relatively long burst when compared to the BATSE 4B revised catalogue \citep{batse}, where the logarithmic mean of the `long burst' distribution ($T_{90}>2$ seconds) is $27.2\pm1.1$ seconds, and only 46 out of the 1234 long bursts had $T_{90}$ greater than 168 seconds. Instrument-to-instrument comparisons of burst durations may be susceptible to differences in detection thresholds \citep{batse_pop} suggesting that the $T_{50}$ duration may be more robust for this comparison. $T_{50}$ for GRB~050117 was $84.796\pm0.010$ seconds, compared with the arithmetic mean $T_{50}$ of the BATSE 4B revised distribution of $16.1\pm1.7$ seconds, indicating that GRB 050117 is in the upper 5\% of the distribution.
 
\begin{figure}[ht]
\figurenum{1}
\epsscale{1.0}
\plotone{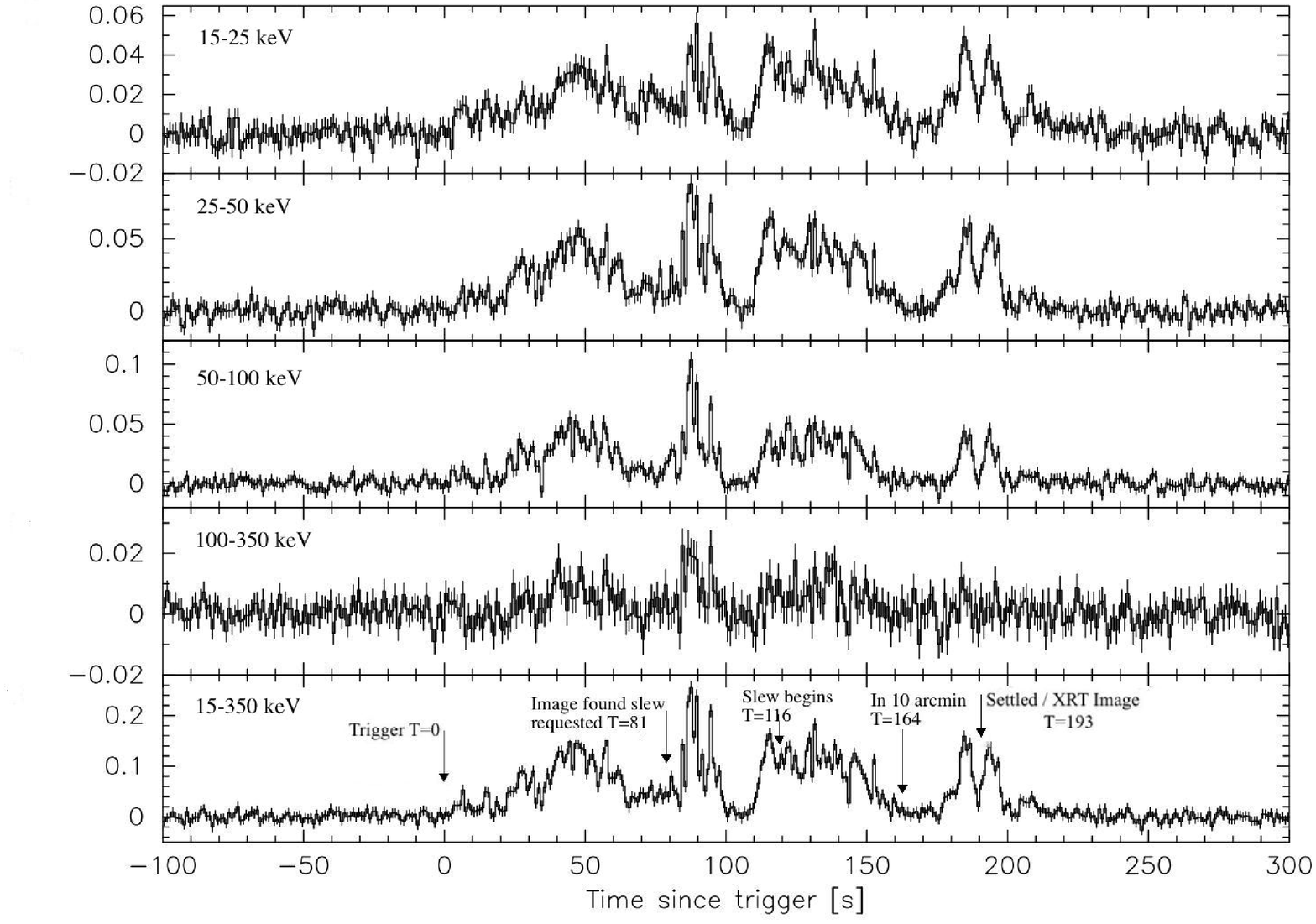}
\caption{BAT energy resolved light curve corrected to on-axis, in counts s$^{-1}$ (fully illuminated detector)$^{-1}$.}
\label{f_energy_res_lc}
\end{figure}

\begin{figure}[ht]
\figurenum{2}
\epsscale{1.0}
\rotate
\plotone{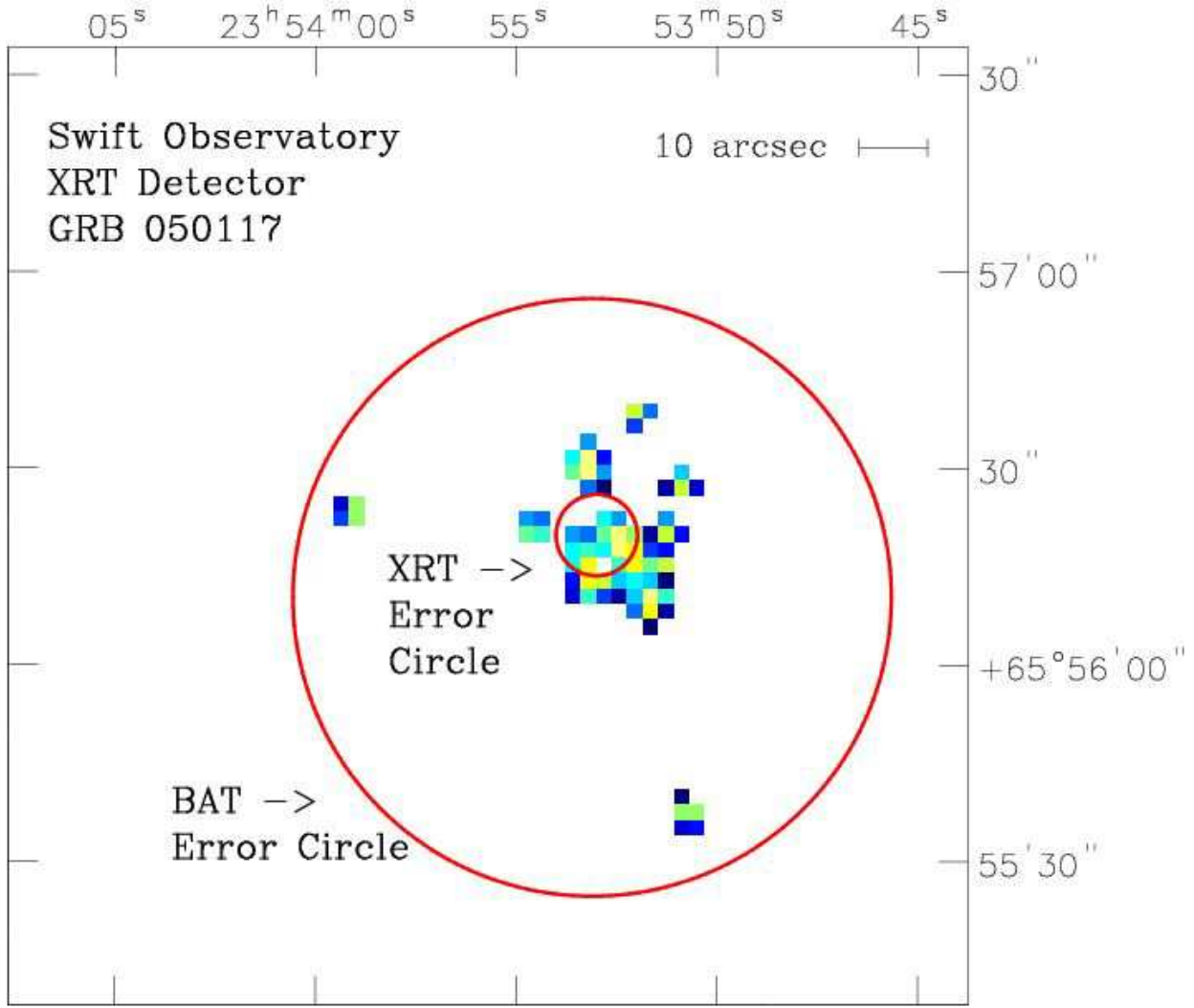}
\caption{XRT Image of GRB 050117, showing the XRT $90\%$ confidence error circle centred on the onboard position, corrected for a non-nominal spacecraft configuration, \xrtposra,\, \xrtposdec, and BAT $90\%$ confidence error circle from ground processing. The observed X-ray flux from this exposure is \xrtfluxerr in the 0.5--10 keV band}.
\label{f_image}
\end{figure}

The BAT background-subtracted lightcurve, for the 15--350 keV energy band, is shown in Figure \ref{f_energy_res_lc} with annotations marking the significant observatory events. Multiple peaks can be seen during the 220 second duration. The peak emission occurred 87.22 seconds after the trigger. By comparing the energy-resolved lightcurves (Figure \ref{f_energy_res_lc}), it can be seen that the initial increase in countrate following the trigger occurred in the soft 15--25 keV band. Subsequently the burst became harder, with more than one third of the total fluence in the 100--150 keV energy band. At the end of the gamma-ray phase, the emission appears to soften.

The BAT time averaged spectra were fitted with two different models; a simple power-law model and a power-law continuum with an exponential cut-off (for the BAT energy band this is equivalent to the lower energy part of the Band function \citep{Band}):

$N_{E}(E)\propto$$E^{\gamma}$$exp(-E(2+\gamma)/E_{peak})$

Where $E_{peak}$ represents the peak energy in the $\nu F_{\nu}$ spectrum and $\gamma$ is the photon index. The fit was significantly improved using a cut-off power law model, with an F-test probability of $3.5\times10^{-6}$. The time-averaged spectral fit over the 15--150 keV energy band gave a photon index of $1.1\pm0.2$ and an $E_{\rm peak}$ of $123^{+50}_{-22}$ keV with a $\chi^2=33.8$ for 56 degrees of freedom (yielding a null hypothesis probability of 0.992).

The burst total energy fluence was $(9.3\pm0.2)\times 10^{-6} $ergs cm$^{-2}$ in the 15--150 keV band in 220 seconds with more than one third of the fluence in the 100--150 keV band, see Table \ref{T4}. The peak photon flux of $2.47\pm0.17$ ph cm$^{-2}$ s$^{-1}$ (integrated for one second from 15--150 keV using the best fit model of a simple power-law), occurred 87.22 seconds after the trigger. 

\begin{figure}[ht]
\figurenum{3}
\epsscale{.50}
\plotone{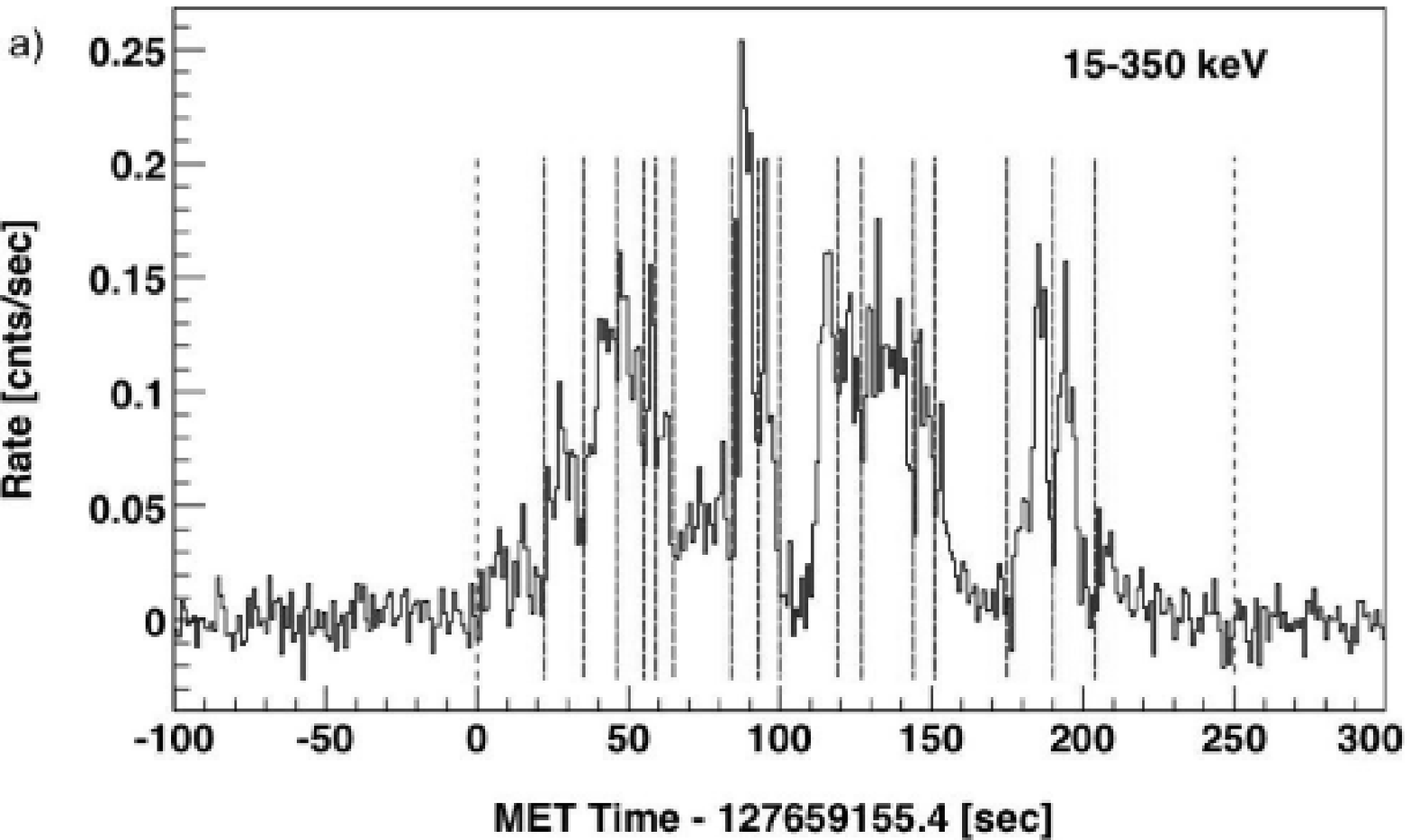}
\plotone{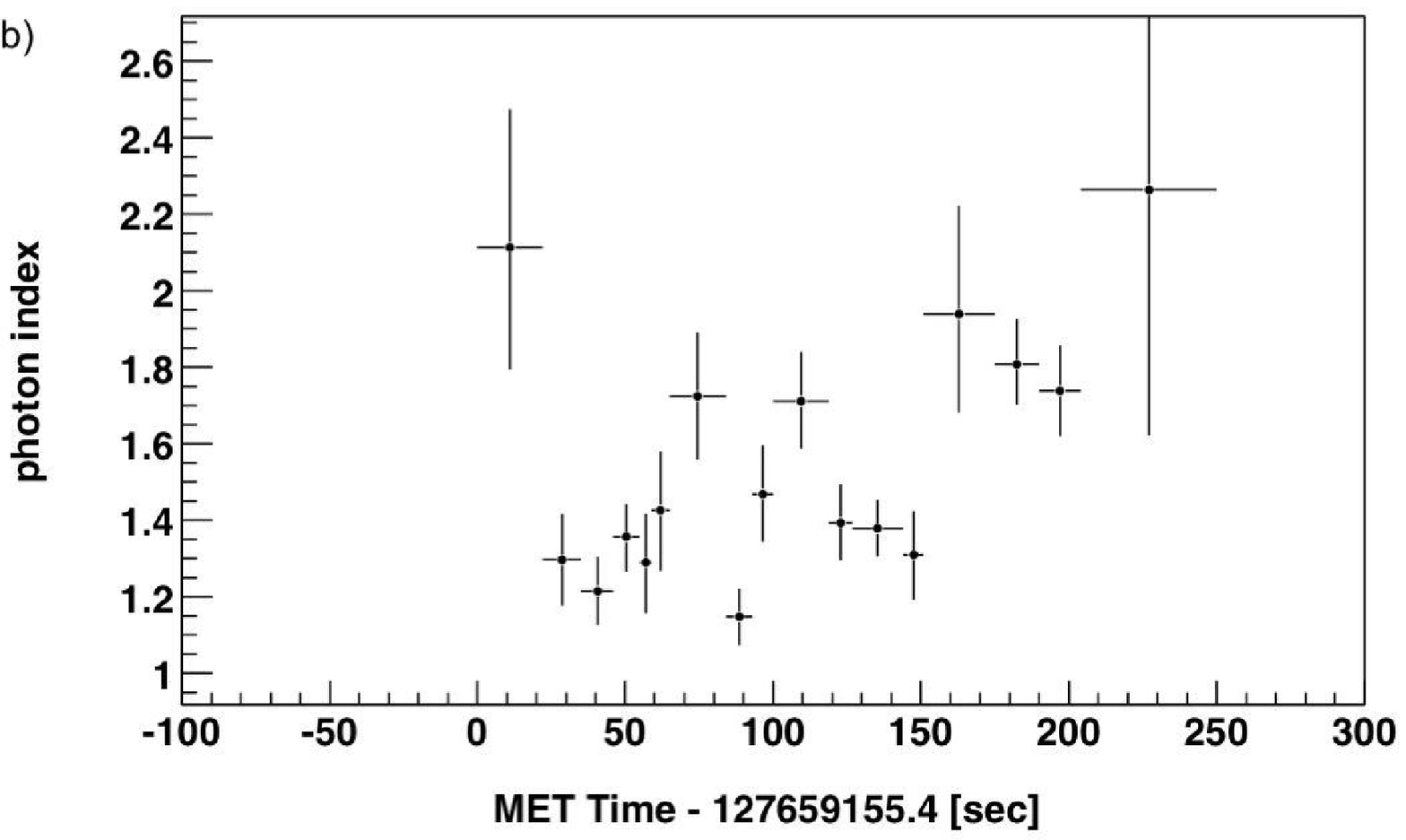}
\plotone{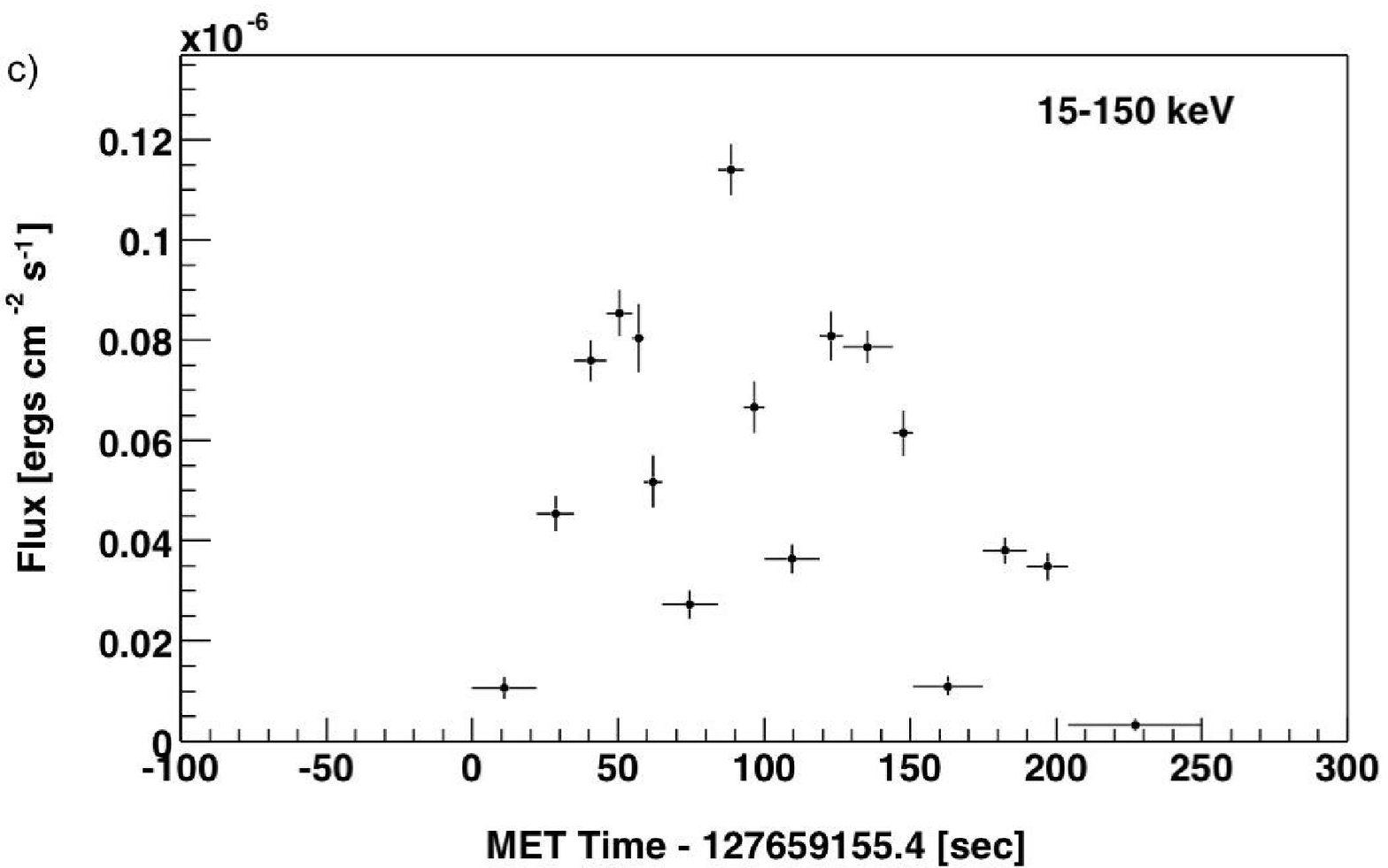}
\caption{BAT time-resolved analysis, where MET 127659155.4 seconds is the trigger time. (a) Background-corrected lightcurve of GRB 050117, in counts $s^{-1}$ (fully illuminated detector)$^{-1}$ corrected to on-axis. Dashed lines define the time bins for the analysis. (b) Photon Index from a simple power law fit from 15--150 keV, corresponding to time bins defined in (a). (c) Energy flux for 15--150 keV.}
\label{f_lightcurve}
\end{figure}

Figure \ref{f_lightcurve} (a) shows the lightcurve obtained from the BAT in the 15--350 keV energy range. The lightcurve is divided into 17 time intervals encompassing each of the multiple peaks and troughs of the lightcurve. A spectrum was extracted for each time interval and fit with a simple power law corrected for the source angle and the respective BAT response. In the short time intervals, the statistics for the time-resolved analysis were not sufficient to warrant the use of a cut-off power law model. The photon index from the simple power-law model spectral fit and the calculated flux for each of the time bins are shown in Figure \ref{f_lightcurve} (b) and (c) respectively. Comparing the photon index with the flux, it can be seen that the spectrum is harder during the peaks of the burst than during the troughs, except for the final peak, where it is comparatively softer. The photon index from the power law spectral fit is $1.8\pm0.1$ during the final peak compared to $\sim1.2$--1.4 during the previous peaks. It is during this final peak that the XRT data were obtained.

\begin{figure}[ht]
\figurenum{4}
\epsscale{1.0}
\plotone{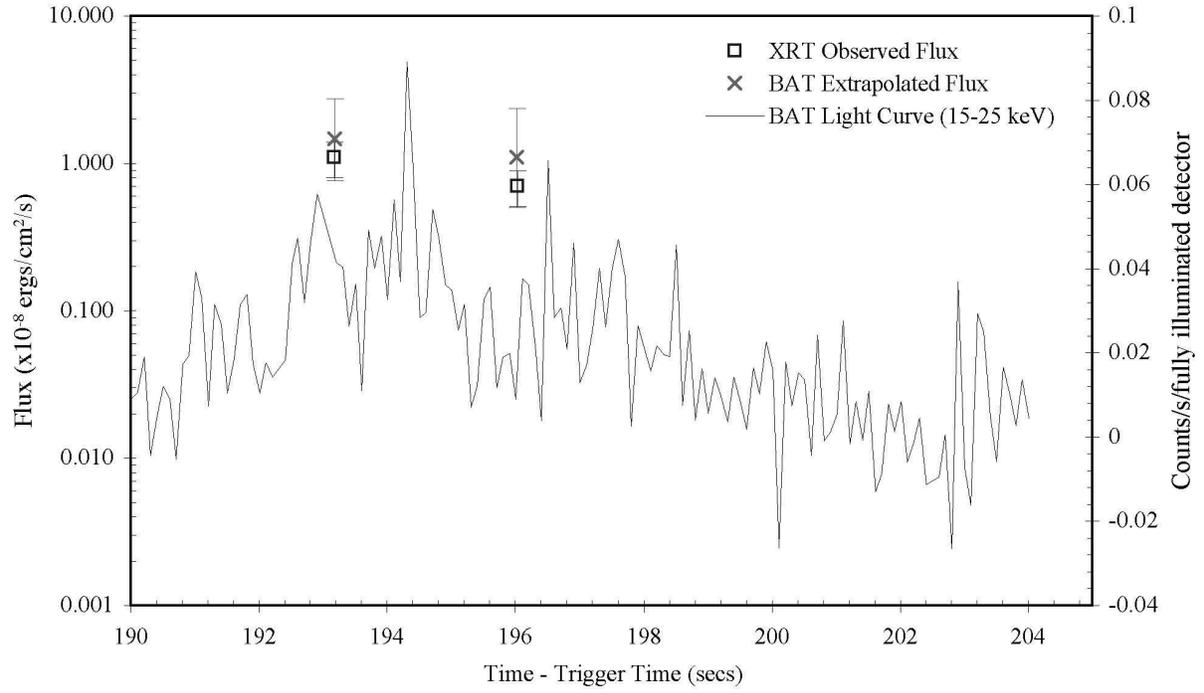}
\caption{ A high resolution plot of the simultaneous X-ray and gamma-ray observation of the prompt emission. The XRT observed flux from 0.5--10 keV, the BAT observed flux from 15--100 keV extrapolated into the XRT 0.5--10 keV band and the last peak of the BAT lightcurve in counts per second per fully illuminated detector. The BAT lightcurve is for the 15--25 keV energy band.}
\label{f_batandxrt}
\end{figure}

\begin{figure}[ht]
\figurenum{5}
\plotone{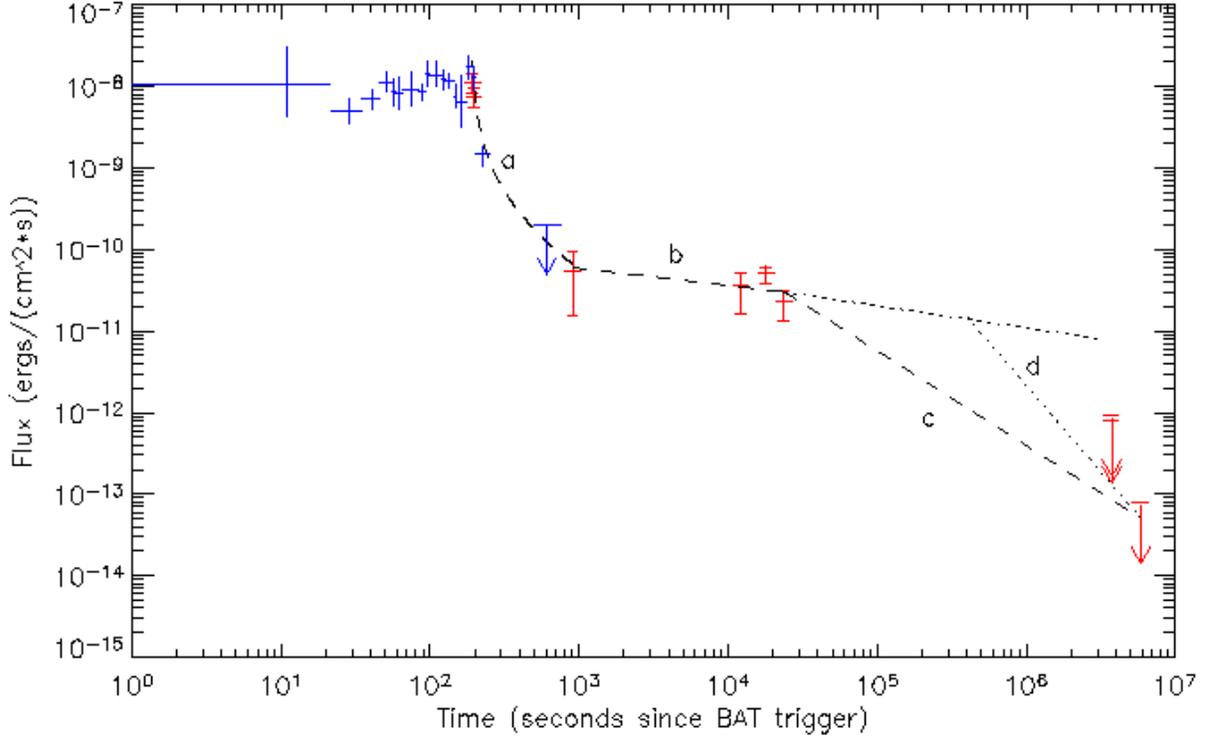}
\caption{GRB 050117 lightcurve (absorbed fluxes): The BAT (blue) lightcurve using the spectral fits for each time bin and extrapolating the 15--100 keV flux into the 0.5--10 keV band (the flux and upper limit for the time intervals $t=204-250$ seconds and $t=300-913$ seconds were calculated using the photon index from $t=190-204$ seconds) and the XRT (red) lightcurve (0.5--10 keV), showing the upper limits for the observations at more than 43 days after the burst; a) A power law fit assuming high latitude emission from the internal shock where $t_{shock}=187$ seconds, $\alpha<-1.2$; b) A power law fit to the afterglow decay with energy input from refreshed shocks assuming a $t_{0}=trigger time$, $\alpha=-0.2\pm0.2$; c) Continuation of the afterglow decay assuming a break in the lightcurve at $t=6.6$ hours, $t_{0}=trigger time$, $\alpha=-1.2$; d) Extrapolation of $\alpha=-2$ from $t=43$ days, showing the latest expected time of the break in the lightcurve at $\sim4.5$ days ($t_{0}=trigger time$).}
\label{f_batxrtlightcurve}
\end{figure}

The BAT spectra below the break energy (15 -- 100 keV) obtained at the time of the two simultaneous XRT measurements ($\pm0.5$ seconds), were fit with a simple power law yielding a photon index of $1.47\pm0.14$ and $1.49\pm0.19$ at the time of the Image and Photodiode Mode data respectively. According to the spectral fit and using the value of the Galactic $N_{\rm H}$=$9\NHunits$ \citep{Nh}, the BAT flux was determined and extrapolated into the XRT 0.5--10 keV band; the results are shown in Figure \ref{f_batandxrt}.  This process was repeated for each of the spectral fits for the 17 time bins shown in Figure \ref{f_lightcurve}; the flux for the 17 BAT time bins extrapolated into the 0.5--10 keV band are shown in Figure \ref{f_batxrtlightcurve}. The photon index obtained for the interval $t=204-250$ seconds is not well constrained (ref figure 3b), and only produces an upper limit for the flux. For this reason, the photon index of $1.7\pm0.1$ obtained for the previous time interval ($t=190-204$ seconds) was used to calculate the flux for $t=204-250$ seconds and to calculate an upper limit for $t=300-913$ seconds. Using the late time photon index of $2.3^{+0.9}_{-0.7}$ increases the flux and upper limit by a factor of 3.5.

Analysis of the BAT data obtained after the end of the prompt emission ($t=300$ seconds) through 913 seconds after the trigger, when the burst became constrained by the Earth for the NFIs, did not detect a source above a $3\sigma$ detection level, corresponding to a flux of $8.5\times 10^{-10} $ergs cm$^{-2}$ s$^{-1}$ between 15--150 keV for a spectrum corresponding to the previous time interval (photon index = 1.5). 

\section{X-ray Observations}
The XRT has several operating modes in order to maximise the science return according to the current observing sequence and the count-rate from the burst. As soon as the observatory settles on a new burst the XRT acquires an image and, if the flux is greater than $\sim14$ mCrab, determines the position of the burst. The position and a postage stamp image centred on the position are telemetered through the Tracking and Data Relay Satellite System (TDRSS). In order to have sufficient full-scale response for very bright bursts, the energy per channel under-samples the spectrum in Image Mode, so that some spectral features are lost. The spectrum is expected to be piled-up, with more than one photon interacting in a pixel.

Following Imaging Mode the XRT switches into Piled-up Photodiode (PuPD) Mode, which provides 0.14 milli-second timing resolution but no positional information and limited spectral information while the burst is bright. According to the source brightness, the XRT automatically switches between PuPD and the following modes; Low-rate Photodiode (LrPD) Mode, where the flux is low enough to obtain spectral information and to update the on-board Photodiode Mode bias level; Windowed Timing (WT) Mode, providing lower timing resolution (1.8 milli-second) than the Photodiode Modes but acquiring a 1--d image and a high resolution spectrum; and Photon-Counting (PC) Mode, a more traditional mode of operation of an X-ray CCD camera. PC Mode is only used when the source is extremely faint and provides limited timing resolution (2.51 seconds), a 2--d image and high resolution spectroscopy. Due to high background rates from the South Atlantic Anomaly (SAA) and the bright Earth limb, the countrate evaluated by the XRT flight software was not sufficiently low to switch into Photon-Counting Mode during the initial observations. The XRT operating modes are described in detail in \cite{modes, modes2} and a summary of the XRT observations of GRB 050117 is provided in Table~\ref{T1}.

GRB 050117 was within the NFI Earth constraint when it was first detected by the BAT. The Observatory slewed when the burst became viewable by the NFIs 117 seconds later, arriving at the burst location 193 seconds after the trigger, by which time the Observatory was in the SAA. The XRT obtained a 0.1 second image of the source immediately after {\it Swift} had settled on the burst and determined a position, which was within the BAT error circle. The Observatory was still in its commissioning phase, therefore the XRT onboard position was corrected on the ground for a non-nominal spacecraft configuration to give an updated position of $\xrtposra$ $\xrtposdec$ \citep{GCN 2955}. The estimated uncertainty in this position is 6.2 arcseconds ($90\%$ confidence).

Following the determination of the GRB position, XRT switched into Piled-up Photodiode (PuPD) Mode at 12:55:51.409 and obtained high resolution timing data for 109 ms before data collection was suspended by the onboard software to avoid flooding the telemetry with proton events from the SAA. There was a period of 2.752 seconds between the Image Mode data and the PuPD data when the instrument switched between modes and no data were obtained. The TDRSS Postage Stamp image, a thresholded image and the PuPD Mode data were obtained by XRT during the later portion of the gamma-ray phase of the burst (Figure \ref{f_energy_res_lc}). The XRT onboard software determined that {\it Swift} had exited the SAA at 13:07:36.24, 900 seconds after the burst detection, and XRT obtained a further 11.15 seconds of Photodiode Mode data prior to the Observatory slewing to a new source due to the approaching Earth constraint.  

The location of GRB 050117 was very close to the {\it Swift} northern observing horizon and therefore was only visible outside of the Earth constraint for a very short period each orbit, the majority of which was while {\it Swift} was in the SAA. No data were telemetered during the second orbit for this reason. Before the third orbit, the XRT automated SAA checking was disabled to permit the XRT to collect data on the burst while in the SAA. By the sixth orbit, the orbital precession had placed the source continuously within the Earth constraint and no further observations were possible. 

\subsection{Image Mode Analysis}
The burst was sufficiently bright at acquisition for the XRT to successfully obtain a refined position from a 0.1 second exposure image. The XRT image of the burst is shown in Figure \ref{f_image}.  

The analysis described here refines the initial flux estimate \citep{GCN 2955}. A source spectrum cannot be extracted from the image, since more than one photon is expected to have interacted in a single pixel. A photo-absorbed power law fit was obtained from the PuPD data immediately following the Image giving a photon index of $2.3\pm0.5$, assuming a Galactic absorbing column of $9\NHunits$ \citep{Nh} and $z=0$ (the details of the spectral fitting are given in Section 3.2). If one assumes the same source spectrum for the Image Mode data, a countrate can be inferred by dividing the total energy from the source in the image by the average energy of the assumed spectrum. Extracting the accumulated charge from the source in the 2 arcmin Postage Stamp Image (defined as $3\sigma$ above the background) and dividing by the average photon energy found from the model, a source countrate of 346 counts s$^{-1}$ is determined. This countrate corresponds to an XRT observed flux of \xrtfluxerr (0.5--10 keV).

\subsection{Photodiode Mode Analysis}
The XRT Photodiode Mode data were complicated by two factors. First, the close proximity of the burst to the Earth limb resulted in an increased low energy background in the data from the bright Earth. This was subtracted by post-processing the data on the ground. The second complicating factor was the increase in background events due to passage through the SAA. A Nickel line is observed in the spectrum, which is an instrumental effect created by interaction of protons in the SAA  with the camera interior.  The higher overall background level produced by the proton events during the SAA meant that the preliminary analysis of the timing and spectral data showed no evidence of emission due to GRB 050117. Analysis of XRT calibration data obtained in the SAA has shown that the overall background contribution to the spectrum during the SAA is proportional to the proton count-rate measured as saturated pixels. This relationship was used to apply background correction to all the data sets on 2005 January 17, following the prompt Image and PuPD data (the low number of proton events in the Image and prompt PuPD data removed the need for such background corrections). Data were only included in the analysis if the number of saturated pixels (due to the SAA) contributed less than 1\% of the total counts yielding a total exposure of 333 seconds. The breakdown of the exposure analysed for each orbit is shown in Table~\ref{T2}. 

After correcting for the bright Earth and subtracting the background attributed to the SAA, the total number of counts with energy greater than $3\sigma$ above the detector noise were calculated. The PuPD Mode data were sufficiently low countrate for one to assume that pile-up effects from the source are negligible and therefore it was possible to extract a countrate. The countrates obtained on 2005 January 17, where $t>900$ seconds, were dominated by the calibration sources and the background contribution from other sources in the field was unknown. For these reasons, two short (approximately 4000 seconds) follow-up observations were performed by {\it Swift} 43 days after the burst. The first was with XRT in LrPD Mode, to determine the counts attributable to the sky and instrument background, without contribution from the source. The second observation was in Photon-Counting Mode to provide two-dimensional imaging. Prompted by very late after-glow detections of other GRBs, an additional 20 ksec observation was made of the GRB 050117 field, primarily in PC Mode, 68 days after the burst. Four more, short exposures were made in LrPD and PC Mode, 71 days after the burst, when GRB 050117 was used as a `fill-in' target for the {\it Swift} observation schedule.

All data obtained in PC Mode, between 2005 March 26--28, were analysed ($>10$ ksec; grades 0--12). From the summed image, there was no detectable source above an upper limit of $3\sigma$ above the background at the position of GRB 050117. The instrumental background is higher in LrPD mode than in PC mode, because the LrPD CCD readout scheme essentially integrates the entire field of view, including the calibration sources, into a single pixel. Based on the evidence that the source had faded below the detectable limit in PC mode, the short exposure LrPD Mode data from 2005 March 29 were used to derive the expected background from the instrument and from the sky for earlier Photodiode Mode observations on 2005 January 17 and March 2. 

For the follow-up observations, where there were a low number of source counts or no detectable source above $3\sigma$ above the background, the expected background counts calculated from the final LrPD observation ($\sim71$ days after the burst) were used to derive $90\%$ confidence intervals and upper limits on the source countrate in accordance with \cite{Kraft}.  

In order to compare the spectral distribution of the prompt X-ray emission with that from the BAT spectral fit, and to determine if there was any obvious spectral evolution between the prompt and follow-up observations, a spectrum was extracted from the prompt PuPD Mode data and a second spectrum was extracted from the data from each orbit following the prompt emission until 6.6 hours after the burst. Channels below 0.3 keV and above 10 keV were ignored and, in addition, channels containing lines from the calibration source and Nickel $K_{\alpha}$ and $K_{\beta}$ from proton interactions with the camera body were also excluded from the analysis.  A photo-absorbed power law model ($z=0$) with an absorption column density fixed to the Galactic value of $N_{\rm H}$=$9\NHunits$ was fit to both data sets.

From the prompt PuPD spectra, a photon index of $2.3\pm0.5$ was determined from a total of 28 counts with a minimum of 9 counts per bin, compared to the BAT photon index of $1.5\pm0.2$. Assuming the same photo-absorbed power-law spectrum with a photon index of 2.3 and a column density of $N_{\rm H}$=$9\NHunits$ we obtain an absorbed flux of $7.3^{+2.4}_{-1.9}\times10^{-9}$ ergs cm$^{-2}$s$^{-1}$ for the 0.5--10 keV energy band. The 109~ms of Photodiode Mode data appear to be a continuation of the prompt emission seen in the 0.1 second Image Mode exposure, at a slightly lower flux. A comparison of the simultaneous BAT and XRT measurements are shown in Figure \ref{f_batandxrt}. The XRT flux measurements obtained in Image Mode and Photodiode Mode of \xrtfluxerr and $7.3^{+2.4}_{-1.9}\times10^{-9}$ ergs cm$^{-2}$s$^{-1}$, respectively, are within $90\%$ confidence of the simultaneous BAT measurements extrapolated into the XRT band, of $1.5^{+1.3}_{-0.7}\times10^{-8}$ ergs cm$^{-2}$s$^{-1}$ and $1.1^{+1.3}_{-0.6}\times10^{-8}$ ergs cm$^{-2}$s$^{-1}$, respectively.

For the follow-up data ($t>900$ seconds), due to the low statistics, the spectra were re-binned with a minimum of only 10 counts per bin. A photon index of $2.0\pm1.1$ was obtained for a photo-absorbed power law model and a column density of $N_{\rm H}$=$9\NHunits$. This spectral fit was used to derive the fluxes or upper limits for all data sets where $t>900$ seconds. The confidence intervals for the fluxes were derived from the confidence intervals for the countrate,  and do not take into account the uncertainty in the spectral fit.

Both the prompt and the follow-up spectral fits indicate a possible low energy contribution, which is not accounted for by a photo-absorbed power law with an absorbing column of the Galactic value. In addition, although the prompt and the follow-up spectral fits are comparable, when the count-rate hardness ratio for the data at $\sim200$ seconds is compared with the hardness ratio for the data obtained between 900 seconds and 6.6 hours, the early data appear to have a harder spectrum. This suggests a spectral evolution from the early data to the later data but there are too few counts to quantify this and this is in contradiction to the photon indices obtained from the spectra. The X-ray fluxes derived from the BAT and the XRT spectral models, for both the early and the follow-up spectra are consistent at the $90 \%$ confidence level between 0.5 and 10 keV.  For this reason, and because the following discussion is not heavily dependent on the spectral analysis, a more rigourous spectral analysis for data with low statistics was not performed. The absorbed fluxes obtained from the XRT spectral fits with $N_{\rm H}$ fixed to the Galactic value (Table~\ref{T2}), are shown with the BAT absorbed fluxes from each of the 17 time intervals extrapolated into the 0.5--10 keV band in Figure \ref{f_batxrtlightcurve}. A smooth transition from the early gamma-ray emission to the X-ray emission is exhibited.  Accounting for the Galactic absorption column, the unabsorbed fluxes are a factor of 2 higher. After the initial detection during the gamma-ray phase of the burst, a faint, slowly decaying afterglow is detected.

On 2005 January 17, a total exposure time of less than 15 seconds was accumulated in Window Timing Mode during which, there was no detectable source above the very high background. It is not possible to correct the bright Earth contribution in the WT data using the same technique as for LrPD Mode, and therefore these data were not included in the analysis.

\section{Follow-up Observations}
Follow-up observations were made at optical and radio wavelengths several hours, and in some cases days, after the burst trigger.  A summary of those observations and the corresponding upper limits taken from GCN Circulars is given in Table~\ref{T3}. 

\begin{figure}[ht]
\figurenum{6}
\epsscale{1.0}
\plotone{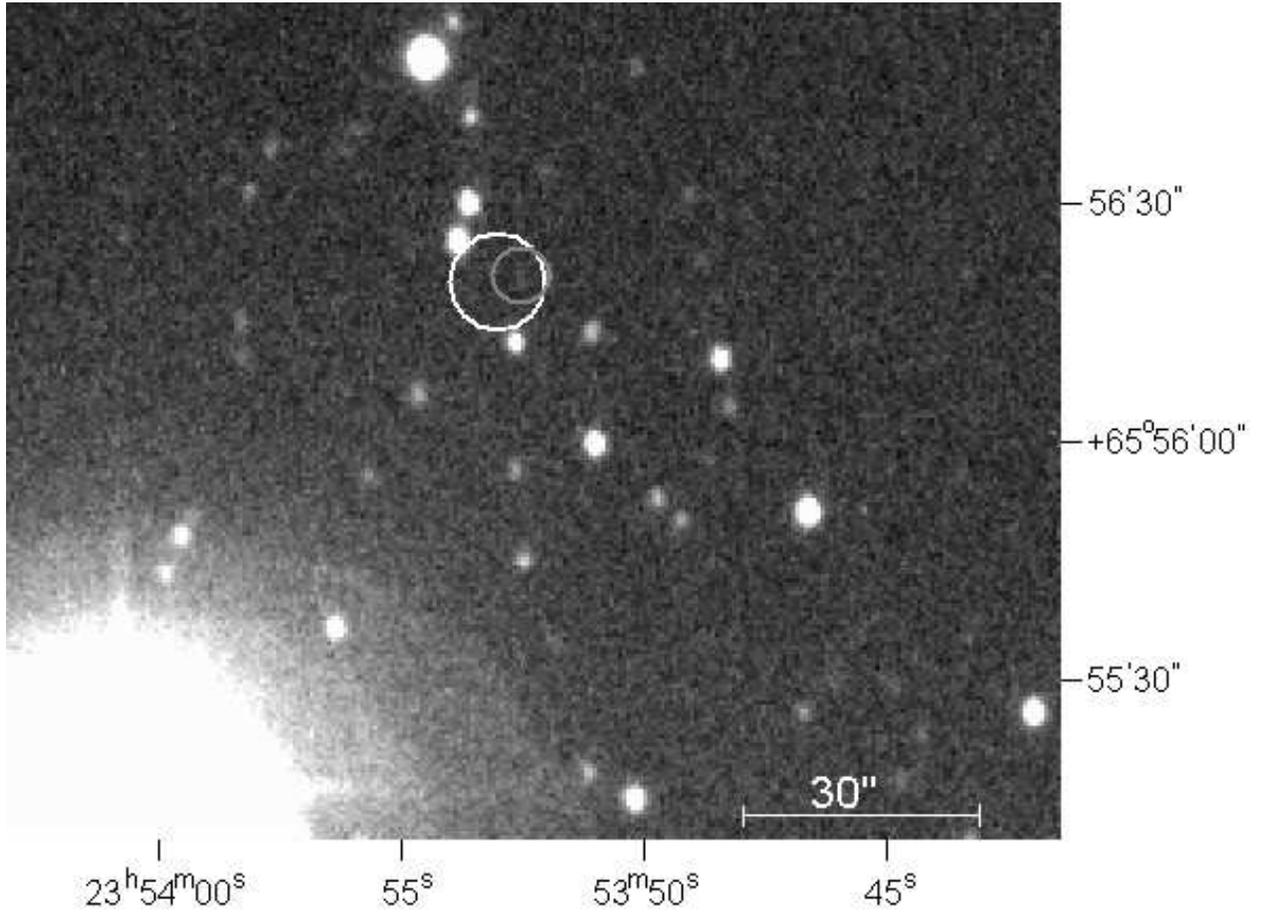}
\caption{Image from the follow-up observation at Mt. Maidanak observatory in R-band. The grey circle denotes the location of the possible source; limiting magnitude $23.9\pm1.3$. The white circle is the XRT error circle.}
\label{f_follow_up}
\end{figure}

A follow-up observation on 2005 January 17 16:58 with the 1.5 m Telescope, Maidanak Astronomical Observatory, identified a source candidate within the XRT  error circle at RA(J2000)=23$^{\rm h}$53$^{\rm m}$52$\fs6$ Dec(J2000)=$+65\degr$56$\arcmin$19\farcs7 (Figure \ref{f_follow_up}). The estimated source brightness, based on a total exposure of 1860 seconds taken in two epochs between 15:49  and 17:22, is $R=23.6\pm0.7$.  Follow-up observations were made on 2005 April 6 with the Crimean Astrophysical Observatory 2.6 m telescope. No object was found down to a limiting magnitude of 23.4 ($3\sigma$) at the position of the source candidate. Due to low signal-to-noise in the January 17 observations and the absence of the underlying  galaxy in the April 6 observations the candidate source  cannot be confirmed as the GRB counterpart. 

The {\it Swift} UV-Optical Telescope was fully activated by the time of the late follow-up observations between 2005 March 26 13:15 and  2005 March 29 22:55. A total exposure of 15817 seconds was obtained with the V-filter. From the co-added images, no source was detected in the XRT error circle down to a limiting magnitude of 21.38 ($5\sigma$).

A lack of optical afterglow can be attributed to three reasons: the afterglow is intrinsically dark, there is dust extinction or the burst has high redshift. \cite{GCN 2956} reported that the Galactic extinction is severe in the direction of the burst ($E_{B-V}$=1.7 or $A_R$=4.6) according to \cite{Schlegel} and therefore the lack of an optical counterpart is not surprising.

\section{Discussion and Conclusions}
The X-ray lightcurve is shown in Table \ref{T2} and Figure \ref{f_batxrtlightcurve}. The X-ray data indicate a decrease in flux of almost three orders of magnitude between the prompt emission at $t=190$ seconds and the afterglow at $t=900$ seconds. An initial steep decay of a similar magnitude over a comparable time period, as measured from the trigger time, has been observed in other {\it Swift} bursts; GRB 050319 \citep{GRB 050319}, GRB 050126 and  GRB 050219a \citep{nature}, but in these cases the gamma-ray emission was over at the time of the X-ray observations. Similarly, the decay becomes significantly flatter over the following $\sim6$ hours and then becomes steeper again sometime later in order to be undetected 43 days after the burst.

The XRT and extrapolated BAT fluxes obtained from the simultaneous XRT and BAT observations (Figure \ref{f_batandxrt}) are within the $90\%$ confidence limits. If we consider that the multiple peaks in the BAT lightcurve are attributed to internal shocks from the collision of the faster expanding shells with slower shells in front, then it is reasonable to assume that the X-ray flux at this time was also produced by an internal shock collision. The minimum expected emission following the peak from the internal shock is the high latitude emission from the curvature effect \citep{naked_burst}. The angular spreading time scale determines the decay timescale, and consequently the width of an internal shock peak, $\delta t$. Therefore, $t_{0}$ of the internal shock is, $t_{shock}=t_{peak}-\delta t=195-8=187$ seconds. Due to the non-detection of gamma-ray flux between $t=300$ seconds and $t=913$ seconds we can assume that the X-ray flux after 900 seconds is dominated by the decaying afterglow and therefore we can use the XRT flux measurement at 900 seconds as an upper limit on the contribution from the decaying internal shock at this time. This provides a constraint on the temporal decay of the internal shock of $\alpha<-1.1$.

Following the end of the prompt emission, the lightcurve enters a shallower decay phase where, for the following 6.3 hours, there is very little decay in flux. The $90\%$ confidence upper limit for the decay index is --0.5, but the best fit to the data is shallower than this; $\alpha=-0.21^{+0.28}_{-0.20}$. A second break in the power law is implied by the steep decay between the data points at 23 ksec and the upper limit at 68 days. In order for the source to be undetected 68 days after the burst, the flux must decay with $\alpha<-1.2$ if the break occurred immediately after the last detection. If the break were later or the flux is significantly less than the upper limit, then the decay could be steeper.

A photo-absorbed power law spectral fit to the prompt emission PuPD data ($t=193$ seconds) using a Galactic $N_{\rm H}$ of $9\NHunits$ yielded a photon index of $2.3\pm0.5$. Fitting the same model to the summed LrPD and PuPD mode data from the afterglow ($t= 900$ seconds -- 6.6 hours) yielded a photon index of $2.0\pm1.1$. For the X-ray data we assume a spectrum of the form F(t,$\nu$)=$(t-t_0)^\alpha$$\nu^\beta$ \citep{grb_ppp}, where $\beta$ is the spectral index and $\beta=1-$ photon index, yielding a spectral index of $-1.3\pm0.5$ and $-1.0\pm1.1$ for the prompt and follow-up observations, respectively.

The following sections discuss the possible theoretical interpretations of the observations.

\subsection{Early Emission (\boldmath{$t<1000$} seconds)}
If the early emission is reviewed in the context of high latitude emission from the internal shock then, taking the photon index of $2.3\pm0.5$ into consideration, where $\alpha=\beta-2$ \citep{naked_burst}, a decay of $-3.3\pm0.5$ would be expected. This is within the constraints of the observation, where $\alpha<-1.1$.

The gamma-ray flux during the last internal shock (Figure \ref{f_lightcurve}) decreases from  $3.5^{+0.3}_{-0.3}\times10^{-8}$ergs cm$^{-2}$ s$^{-1}$ at 197 seconds to $3.3^{+1.4}_{-1.2}\times10^{-9}$ergs cm$^{-2}$ s$^{-1}$ at 227 seconds. Using $t_{0}=t_{shock}=187$ seconds, indicates a gamma-ray decay index of --1.7, or --2.1 taking the extremes of the $90\%$ confidence limits. This is less steep than one would expect from high latitude emission where, for a BAT photon index of $1.5\pm0.2$, $\alpha$= $\beta-2=-2.5$. To satisfy a decay of --2.5, $t_{0}$ would have to be further from the observed peak at 178 seconds. The observed flatter decay implies that there is an additional internal shock contribution of a lesser intensity. However, the decay is steep enough to be within the constraint determined in the X-ray regime. Due to the XRT and extrapolated BAT fluxes being of the same magnitude, one can assume that the X-ray flux during the prompt phase is dominated by the emission from the internal shock and any afterglow contribution, which one may expect at this time, cannot be detected. The expected afterglow decay may be in progress between the last internal shock and the flattening of the light curve at 900 seconds, but this cannot be concluded from these data.

Alternatively, if we fit a power law with $t_0=trigger time$ to the two prompt measurements and the measurement at 900 seconds, we obtain a decay of $\alpha=-3.5^{+0.6}_{-0.8}$. Because there are no measurements between 190 and 900 seconds, this is an upper limit on the decay index and the decay could be steeper. If we consider that the early X-ray flux is from the afterglow, we see that the rapid decay is inconsistent with the spherical blast wave model, but the observed $\alpha$/$\beta$ relation maybe consistent with the evolution of a jet \citep{jet_angle}. From the equations in Table 1 of \cite{jet_angle}  if the X-ray frequency is below the cooling frequency ($\nu<\nu_{c}$), for a spectral index $\beta=-1.3$, we would expect a temporal index of $\alpha=2\beta-1=-3.6$ indicating an electron energy distribution index of $p\sim3.6$. The expected decay from the evolution of a jet is within $90\%$ confidence limits of the observed decay, but the decay of the later emission ($t>1000$ seconds) is much shallower than that which is expected after a jet-break. If instead we assume $\nu>\nu_{c}$, we find the observed $\alpha$ is also inconsistent with an expected decay of -2.6.

\subsection{Late Emission (\boldmath {$t>1000$} seconds)} 
The shallow decay, $\alpha=-0.21^{+0.28}_{-0.20}$, may be explained by refreshed shocks, for which there are three possible mechanisms. The first is that the Lorentz factor has a power law distribution, so that the slow shells pile up onto the decelerated blast wave to energise the shell \citep{rees_mes, pan_mes_rees, sari_mes}. Second, the central engine is continuously injecting energy with a (possibly) reduced rate \citep{zha_mes_1} and finally, there may be collisions between the late injected shells with the decelerated fireball \citep{ku_pir, zha_mes_2}. In such a model one could expect undetected lightcurve bumps if the Lorentz factor distribution or the outflow rate is very discontinuous.

This burst was long and multi-peaked, and therefore if the later shells were slow moving with a modest Lorentz factor, energy would be injected into the afterglow as each shell collides into the external medium. This would cause re-brightening super-imposed on the nominal afterglow decay and could explain the flatter than expected decay between 900 seconds and 6.6 hours. The lightcurve is not well sampled and so bumps which may be expected from re-brightening cannot be discerned from the lightcurve. The flattening of the lightcurve to a decay of --0.2 during this period and the observed $\alpha$/$\beta$ relation does not satisfy a simple synchrotron shock model outlined in Table 1 of \cite{grb_ppp}. The break in the lightcurve could be interpreted within the context of a simple afterglow model, satisfying the requirements of Table 1 of \cite{jet_angle}, provided one can assume a small electron index of 1.3 if $\nu<\nu_{c}$ or 0.9 if $\nu>\nu_{c}$, and the large error on the spectral index is taken into account ($\beta=-1.0\pm1.0$). A mechanism for producing such flat electron indices has been discussed by \cite{small_p}.

If the steepening is treated as the classical jet break due to the deceleration of the fireball, we can set limits on the width of the jet using the relation between jet break time and jet opening angle from \cite{jet_angle}:

$\theta_{jet}=10\times(t_{jet}(hrs)/6.2)^{3/8}$

If we assume that the earliest the jet break occurs is at the next to last $3\sigma$ detection, five hours after the burst trigger, we calculate a jet angle of $\theta_{jet}\sim3$ degrees. Alternatively, we can assume that the jet break occurs between the last $3\sigma$ detection at 6.6 hours and the upper limits at 43 days. If we assume a power law decay index of --2 after the break (which is typical of post-jet break decay indices), and extrapolate back from the upper limit at 43 days, we find that the break would have to occur at $t\sim4.5$ days. This implies a jet opening angle of $\sim10$ degrees. Jet angles of 3--10 degrees are consistent with typical GRB jet opening angles \citep{frail}. We note, however, that the slope of the lightcurve between 900 seconds and five hours ($\alpha=-0.2$) is significantly flatter than that of the typical pre-jet break decay slope ($\alpha=-1$). 

In the refreshed shock scenario, we assume that the afterglow due to ejected material colliding with the ISM began at some early time while the prompt emission was still in progress for the observer. If the afterglow were sufficiently bright during the XRT observations at $t=193$ seconds and $t=196$ seconds, we would expect to see the afterglow emission super-imposed on top of the internal shock emission, and thus a difference between the X-ray and extrapolated gamma-ray fluxes. We can estimate the expected flux contribution from the afterglow by assuming that the afterglow decays with a typical slope of $\alpha=-1$ prior to the onset of the refreshed shocks. If we project a power-law with $\alpha=-1$  back from $t=900$ seconds, we find that the expected contribution from the afterglow at t=196 seconds is  $\sim3\times10^{-10}$ergs cm$^{-2}$ s$^{-1}$. This is an order of magnitude below the uncertainty in X-ray flux measurements at that time. Thus, any addition flux contribution from the afterglow in the XRT band pass at $t=193$ seconds and $t=196$ seconds is within the uncertainty of the flux measurement and thus the presence of the afterglow at that time is not inconsistent with the data. 

A steepening of the lightcurve is expected when all of the shocks have collided with the decelerating blast wave and the nominal decay is resumed. This is supported by the data where, in order for the afterglow to be undetected 68 days after the burst,  the lightcurve must steepen to $<-1.2$ at some time after 6.6 hours.

\subsection{Summary}
The early X-ray lightcurve between  $t=193$ seconds and $t=900$ seconds is well explained by a simple internal shock mechanism, beginning at $t_{0}=t_{shock}=187$ seconds after the burst trigger time and decaying with high latitude emission, $\alpha<-1.2$. The afterglow due to the collision with the ambient medium may have started while the internal shock emission was still in progress, although the sample rate is too low to confirm this. The emission from the afterglow appears to be further enhanced by the additional input of energy from lagging shells of ejected material. The refreshed shock energy injection continues until at least five hours after the burst. Some time between five hours and 4.5 days after the burst trigger, the refreshed shocks ceased and the lightcurve turned over to a steeper decay rate of $\alpha<-1.2$ corresponding to the expected afterglow decay, thus the burst was below the XRT detection threshold at $t=68$ days.
 
To date, there have been no other observations by {\it Swift} with simultaneous gamma-ray and X-ray detections. The observations of GRB 050117 demonstrate the unique capability of {\it Swift} to observe both the burst and the afterglow in the X-ray regime.

\acknowledgments
This work is supported at Penn State by NASA contract NAS5-00136; at the University of Leicester by the Particle Physics and Astronomy
Research Council on grant number PPA/Z/S/2003/00507; and at OAB by funding from ASI on grant number I/R/039/04.  We gratefully acknowledge the contributions of dozens of members of the {\it Swift} team at PSU, University of Leicester, OAB, GSFC, ASDC and our subcontractors, who helped make this Observatory possible and to the Flight Operations Team for their dedication and support.

\clearpage

\begin{deluxetable}{ccccc}
\tablecaption{GRB 050117: Energy Fluence and Peak Photon Flux measured by the BAT.
\label{T4}}
\tabletypesize{\normalsize}
\tablecolumns{5}
\tablewidth{0pt}
\tablehead{
\colhead{Energy Band} &
\colhead{Energy Fluence\tablenotemark{a,}\tablenotemark{c}}&
\colhead{Peak Photon Flux\tablenotemark{b,}\tablenotemark{c}}\\
\colhead{(keV)} &
\colhead{ (ergs cm$^{-2}$)} &
\colhead{(photons cm$^{-2}$ s$^{-1}$)} 
}
\startdata
15--25     & $(1.17\pm0.04)\times10^{-6}$   & $(5.45\pm0.78)\times10^{-1}$\\
25--50     & $(2.40\pm0.05)\times10^{-6}$   & $(7.43\pm0.64)\times10^{-1}$\\
50--100   & $(3.44\pm0.08)\times10^{-6}$    & $(7.46\pm0.54)\times10^{-1}$\\
100--150 &$(2.33\pm0.10)\times10^{-6}$    &$(4.38\pm0.50)\times10^{-1}$\\
{\nodata} &{\nodata} 					&{\nodata}  \\
15--150   &$(9.33\pm0.20)\times10^{-6}$   &$(2.47\pm0.17)$
\enddata
\tablenotetext{a} {Cut-off power law model}
\tablenotetext{b} {Simple power law fit}
\tablenotetext{c}{90\% error}
\end{deluxetable}
\clearpage

\begin{deluxetable}{ccclll}
\tablecaption{A Summary of XRT Observations of GRB 050117\label{T1}}
\tabletypesize{\normalsize}
\tablecolumns{6}
\tablewidth{0pt}
\tablehead{
\colhead{Orbit \#\tablenotemark{a}} & 
\colhead{Start Time} & 
\colhead{Stop Time} & 
\colhead{Live-time} & 
\colhead{Time Since BAT} & 
\colhead{XRT Mode}\\
\colhead{or date} &
\colhead{(UT)} &
\colhead{(UT)} &
\colhead{(Seconds)} &
\colhead{Trigger (Seconds)} &
\colhead{\nodata}
}
\startdata
1\tablenotemark{b} &    12:55:49.24	& 12:55:49.34 &	0.1          &	193.2 secs   &	Image\\
1\tablenotemark{b} &	12:55:52.42	& 12:55:52:53	&       0.109     &196.0 secs   &	PuPD\\
1\tablenotemark{b} &	13:07:36.2	& 13:07:47.4&	11.2  &900.2 secs   &	LrPD/PuPD\\
3\tablenotemark{b} &	16:11:06.9	& 16:19:16.1	&       485	&3.3 hrs 	& LrPD/PuPD\\
4\tablenotemark{b} &	17:49:19.2	& 17:54:42.0	&       331	& 5.0 hrs 	& LrPD/WT/PuPD\\
5\tablenotemark{b} &	19:27:53.7	& 19:29:28.5  &     119 & 6.6 hrs &	LrPD/WT\\
2005 March 02 &  00:12:43.8 & 05:27:45.3  & 7711 & 43.5 days & LrPD/PC\\
2005 March 26 & 13:15:38.8 & 05:35:57.8\tablenotemark{c} &  10 346 & 68.0 days& PC\\
2005 March 29 &  21:12:18.2 & 21:15:44.1 & 197 & 70.6 days & LrPD\\

\enddata
\tablenotetext{a}{Source visibility during orbit 2 was entirely in the SAA and before SAA checking was disabled}
\tablenotetext{b}{2005 January 17}
\tablenotetext{c}{2005 March 28}
\end{deluxetable}
\clearpage

\begin{deluxetable}{ccccccc}
\tablecaption{XRT Absorbed Flux Measurements ($90\%$ Confidence Intervals) of GRB 050117 in the 0.5--10 keV band\label{T2}}
\tabletypesize{\normalsize}
\tablecolumns{7}
\tablewidth{0pt}
\tablehead{
\colhead{Useful Exposure} & 
\colhead{Start Time} & 
\colhead{Lower Limit} &
\colhead{XRT Flux} &
\colhead{Upper Limit} \\
\colhead{(Seconds)} &
\colhead{(since BAT trigger)} &
\colhead{(0.5--10 keV)} &
\colhead{(0.5--10 keV)} &
\colhead{(0.5--10 keV)}\\
\colhead {} &
\colhead {(Seconds)} &
\colhead{(ergs cm$^{-2}$ s$^{-1}$)} &
\colhead{(ergs cm$^{-2}$ s$^{-1}$)} &
\colhead{(ergs cm$^{-2}$ s$^{-1}$)}
}
\startdata
0.100     &	193.2 secs   &	$0.8\times10^{-8}$ &	$1.1\times10^{-8}$ & $1.4\times10^{-8}$ \\
0.109     &196.0 secs   &	$5.4\times10^{-9}$ &	$7.3\times10^{-9}$ & $9.7\times10^{-9}$\\
11.1  & 900.2 secs   &	$1.5\times10^{-11}$ &	$5.5\times10^{-11}$ &  $9.5\times10^{-11}$ \\
41.3	&3 hrs 18 min	&	$1.6\times10^{-11}$ & $3.6\times10^{-11}$  & $5.1\times10^{-11}$\\
173.5	& 4 hrs 57 mins	&	$3.8\times10^{-11}$ & $5.1\times10^{-11}$  & $6.1\times10^{-11}$ \\
107.4 & 6 hrs 35 mins &	$1.3\times10^{-11}$ &	$2.3\times10^{-11}$& $3.1\times10^{-11}$\\
905.8 & 43.47 days & \nodata & \nodata & $6.9\times10^{-13}$\tablenotemark{a}\\
905.8 & 43.53 days & \nodata & \nodata & $8.3\times10^{-13}$\tablenotemark{a}\\
10 346 & 68.02 days& \nodata &\nodata&$3.2\times10^{-14}$\tablenotemark{a}\\
196.5 & 70.56 days & \nodata & \nodata & $3.4\times10^{-12}$\tablenotemark{a}\\
\enddata
\tablenotetext{a}{no detection, $90\%$ upper limit}
\end{deluxetable}
\clearpage

\begin{deluxetable}{lclcl}
\tablecaption{Ground-based Follow-up Observations of GRB 050117\label{T3}}
\tabletypesize{\normalsize}
\rotate
\tablecolumns{5}
\tablewidth{0pt}
\tablehead{
\colhead{Telescope} &
\colhead{Start Time} &
\colhead{Filter} &
\colhead{Duration} &
\colhead{Upper Limit}\\
\colhead{} &
\colhead{(UT)}&
\colhead{} &
\colhead{} &
\colhead{}
}
\startdata
14ÕÕ ART Osaka	& 2005 January 17 14:05	& Ic band   & 10 x 60 secs & $>17.5$ \\
\citep{GCN 2956}&  \nodata & \nodata & \nodata & \nodata\\

MASTER Robotic Telescope & 	2005 January 17 14:58	& Unfiltered & 50 x 45 secs & $>19.0$\\
Moscow \citep{GCN 2953}   & \nodata & \nodata & \nodata  \nodata \\
\citep{GCN 2954} & \nodata & \nodata & \nodata & \nodata\\

1.5 m, Maidanak	  &        2005 January 17 15:49	&R band  & 1860 secs  & $23.6\pm0.7$ \tablenotemark{a}\\
\citep{GCN 2958} & \nodata &  \nodata &  \nodata &  \nodata\\

1.5 m, Granada	  &       2005 January 17 18:30	& I band & 6 x 300 & $>20.5$\\
\citep{GCN 2957}& \nodata & \nodata & \nodata &\nodata\\

P200/WIRC Palomar Hale& 	2005 January 18 03:30 & 	Ks band & 32 mins & $>18$\\
 \citep{GCN 2960} \citep{Berger} & \nodata & \nodata & \nodata &\nodata\\

VLA \citep{GCN 2963}	&     2005 January 19 01:55	&  8.46 GHz & \nodata & $<130$ $\mu$Jy \\
VLA	\citep{GCN 2980} &          2005 January 24 03:21	  & 8.46 GHz & \nodata & $< 56$ $\mu$Jy\\

2.6 m, CrAO	  &        2005 April 6 17:59	&R band  & 3060 secs  & $>23.4$\\
\enddata
\tablenotetext{a} {possible source identified} 
\end{deluxetable}
\clearpage

\clearpage

\clearpage

\clearpage

\clearpage

\clearpage

\clearpage

\end{document}